\documentclass[12pt,twoside]{article}

\usepackage{epsf}
\usepackage{wrapfig}
\usepackage{graphicx}
\usepackage{amssymb}
\usepackage{cmp2e}

\title[\mathsurround=0pt%
Studying spin-$\frac{1}{2}$ $XY$ chains]
{\mathversion{bold}\mathsurround=0pt%
There is life in the old horse yet or what else we can learn
studying spin-$\frac{1}{2}$ $XY$ chains}
\author[O.Derzhko]%
{O.Derzhko\refaddr{label1,label2}}
\addresses{
\addr{label1}
Institute for Condensed Matter Physics\\
of the National Academy of Sciences of Ukraine,\\
1 Svientsitskii Str., 79011 Lviv, Ukraine
\addr{label2}
Chair of Theoretical Physics,\\
Ivan Franko National University of Lviv,\\
12 Drahomanov Street, 79005 Lviv, Ukraine
}

\date{Received September 2, 2002}

\begin{document}
\setcounter{page}{729}
\maketitle

\begin{abstract}
\mathsurround=0pt%
We review some recent results on statistical mechanics of the
one-dimens\-ional spin-$\frac{1}{2}$ $XY$ systems paying special
attention to the dynamic and thermodynamic properties of the
models with Dzyaloshinskii-Moriya interaction, correlated
disorder, and regularly alternating Hamiltonian parameters.
\pacs 75.10.-b
\end{abstract}

\section{Introductory remarks}

One-dimensional spin-$\frac{1}{2}$ $XY$ model in a transverse field
defined by the Hamiltonian
\begin{eqnarray}
\label{01}
H=\sum_n \Omega s_n^z
+\sum_n \left(J^x s_n^xs_{n+1}^x + J^y s_n^ys_{n+1}^y\right)
\end{eqnarray}
is known as the simplest quantum many-body system for which many
statistical mechanical calculations can be performed exactly,
i.e., without making any simplifying approximations. For more than
forty years this model has been a standard testing-ground for
checking various conjectures or new calculation schemes and
approaches in statistical mechanics and condensed matter physics.
The aim of the present paper is to elucidate some recent results
derived for spin-$\frac{1}{2}$ $XY$ chains and to foresee some
further problems which are attractive to study. The interest in
spin-$\frac{1}{2}$ $XY$ chains may be enforced nowadays because of
the progress in material sciences (see, for example, the recent
report on Cs$_2$CoCl$_4$, the compound which is a good realization
of the famous spin-$\frac{1}{2}$ isotropic $XY$ chain \cite{001}).

\section{Dynamic properties in fermionic picture}

Spin-$\frac{1}{2}$ $XY$ chains contain a hidden symmetry which was
discovered by applying the Jordan-Wigner transformation: the
system of interacting spins (\ref{01}) can be described in terms
of noninteracting spinless fermions (E.Lieb, T.Schultz, D.Mattis).
As a result many (although by no means all) statistical mechanical
calculations can be performed rigorously. As an example of
notorious problems in the statistical mechanics of
spin-$\frac{1}{2}$ $XY$ chains we may mention the analysis of the
time-dependent correlation functions of $x$ ($y$) spin components
$\langle s_n^x(t)s_{n+m}^x\rangle$,
$
s_n^x(t)=\exp({\mbox{i}}Ht)s_n^x\exp(-{\mbox{i}}Ht)
$%
,
$
\langle(\ldots)\rangle ={\mbox{Tr}}(\exp(-\beta
H)(\ldots))/{\mbox{Tr}}(\exp(-\beta H))
$%
.
Since the relation
between the $x$ spin component attached to a certain site and the
on-site creation and annihilation operators of fermions is
nonlocal and involves the occupation-number operators of fermions
at all previous sites, the problem of applying
the Wick-Bloch-de~Dominicis theorem
to a product of a huge number of multipliers
arises. The result can be written compactly as the Pfaffian of an
antisymmetric matrix (generally speaking of huge sizes)
constructed from the elementary contractions and hence a further
analytical analysis becomes not simple. The problem has been
solved to some extent by elaborating the numerical schemes for
computation of Pfaffians \cite{002,003,004,005,006}. The
numerically derived%
\footnote {The numerical approach is not
restricted to the uniform chains and can be easily applied to the
nonuniform chains in which the Hamiltonian parameters vary
regularly along the chain with a finite period or are random
variables with a given probability distribution \cite{007}.}
results for $xx$ ($yy$, $xy$, $yx$) dynamics
supplemented by the analytical results for $zz$ dynamics
\cite{008}
permit to work out the theory of dynamic properties
of spin-$\frac{1}{2}$ $XY$ chains
in the fermionic picture
(G.M\"{u}ller with coworkers)
and thus to explain
the peculiarities of responses of the spin system
to small external perturbations
\cite{008,009}.

\begin{figure}[tb]
\centerline{\includegraphics[width=\textwidth]{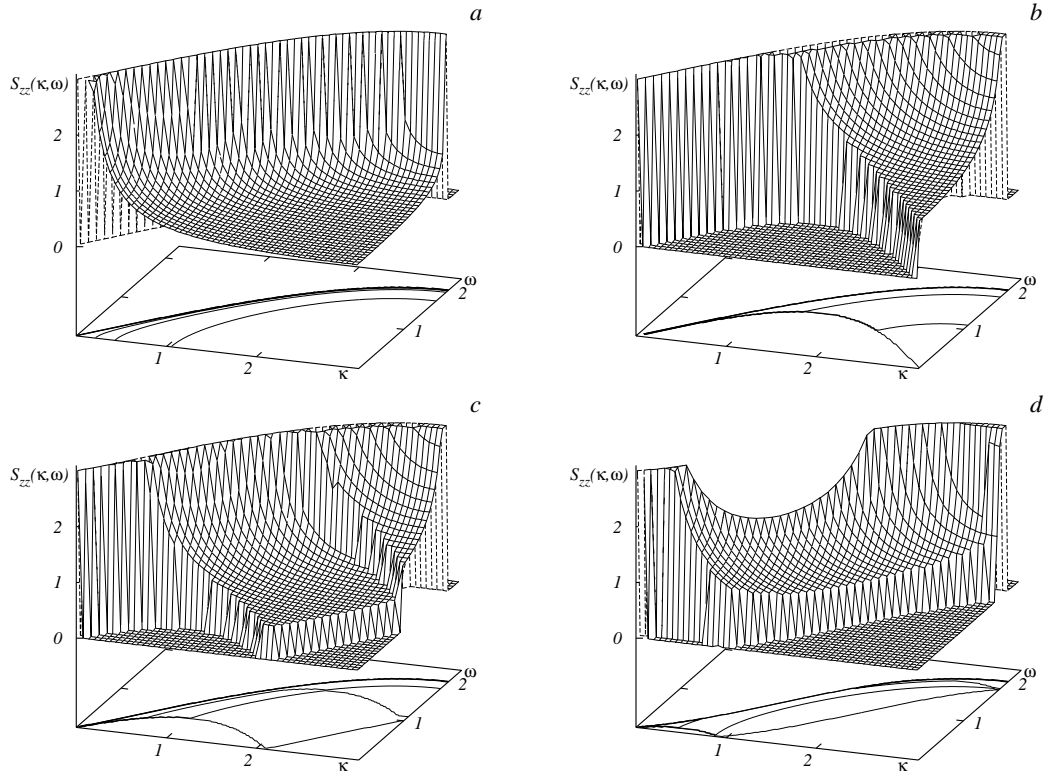}}
\caption{$S_{zz}(\kappa,\omega)$ for the isotropic $XY$ chain
in a transverse field ($J=1$)
at infinite temperature $\beta=0$
(a)
and
zero temperature $\beta\to \infty$
(b: $\Omega=0$, c: $\Omega=0.5$, d: $\Omega=0.9$).}
\label{fig01}
\end{figure}

Let us sketch briefly
the linear response theory of the spin chain in the fermionic language
considering for simplicity the isotropic $XY$ model
($J^x=J^y=J$).
The $zz$ dynamic structure factor for this model
which is given by
\begin{eqnarray}
\label{02}
S_{zz}(\kappa,\omega)
=\int_{-\pi}^{\pi}
{\mbox{d}}\kappa_1 n_{\kappa_1}\left(1-n_{\kappa_1-\kappa}\right)
\delta\left(\omega+\Lambda_{\kappa_1}-\Lambda_{\kappa_1-\kappa}\right),
\\
\Lambda_\kappa=\Omega + J\cos\kappa,
\;\;\;
n_\kappa=\frac{1}{1+\exp(\beta\Lambda_\kappa)}
\nonumber
\end{eqnarray}
suggests the following interpretation \cite{008,010}. Consider at
first the high-temperature limit $\beta\to 0$ when $n_\kappa\to
\frac{1}{2}$ and hence $S_{zz}(\kappa,\omega)$ (\ref{02}) becomes
independent of $\Omega$. Applying the infinitesimally small
external field (directed along $z$ axis) characterized by the wave
vector $\kappa$ and frequency $\omega$ we observe that the
responsive magnetization (directed along $z$ axis) is determined
by generation of the two fermions with energies
$\Lambda_{\kappa_1}$ and $\Lambda_{\kappa_2}$ under the
restrictions $\kappa=\kappa_1-\kappa_2$ and
$\omega=-\Lambda_{\kappa_1}+\Lambda_{\kappa_2}
=2J\sin{\frac{\kappa}{2}}\sin\left(\kappa_1-\frac{\kappa}{2}\right)$.
The ``dummy'' wave vector $\kappa_1$ in (\ref{02}) varies within
the region $-\pi\leqslant\kappa_1\leqslant\pi$. As a result, such an
experimental probe ``measures'' a continuum of the two-fermion
excitations in the $\kappa$-$\omega$ plane. The upper boundary of
the two-fermion continuum is given by $\omega=2\vert
J\sin\frac{\kappa}{2}\vert$ and $S_{zz}(\kappa,\omega)$ exhibits
divergence along this line as it follows from (\ref{02}). At low
temperatures $\beta\to\infty$, the Fermi factors in (\ref{02})
come into play and $S_{zz}(\kappa,\omega)$ becomes dependent on
$\Omega$. The additional conditions $\Lambda_{\kappa_1}<0$ and
$\Lambda_{\kappa_2}>0$ lead to the appearance of the lower
boundary (for example, $\omega=\vert J\sin\kappa\vert$ if
$\Omega=0$) at which a finite value of $S_{zz}(\kappa,\omega)$
jumps to zero. The transverse field (which plays a role of the
chemical potential in the fermionic picture) effects the lower
boundary of the two-fermion continuum in the $\kappa$-$\omega$
plane and the redistribution of the values of
$S_{zz}(\kappa,\omega)$ in the $\kappa$-$\omega$ plane at low
temperatures. For $\vert\Omega\vert>\vert J\vert$
$S_{zz}(\kappa,\omega)$ vanishes everywhere in the
$\kappa$-$\omega$ plane. The above-said can be seen in figure~1
where some typical results
illustrating $zz$ dynamics are reported.

Contrary to the $zz$ dynamics the $xx$ dynamics is more involved:
we do not know the explicit expression for $S_{xx}(\kappa,\omega)$
similar to (\ref{02}).
Equation~(\ref{02}) arises after computation of the average
of the product of four Fermi operators
\begin{eqnarray}
\left\langle
\left(1-2c_n^+(t)c_n(t)\right)
\left(1-2c_{n+m}^+c_{n+m}\right)
\right\rangle
\nonumber
\end{eqnarray}
that is obviously the two-fermion quantity.
The $xx$ dynamic structure factor contains the averages like
\begin{eqnarray}
\left\langle
\left(1-2c_1^+(t)c_1(t)\right)
\ldots
\left(1-2c_{n-1}^+(t)c_{n-1}(t)\right)
\left(c_n^+(t)+c_n(t)\right)
\right.
\nonumber\\
\left. \times \left(1-2c_1^+c_1\right) \ldots
\left(1-2c_{n+m-1}^+c_{n+m-1}\right)
\left(c_{n+m}^++c_{n+m}\right) \right\rangle \nonumber
\end{eqnarray}
and thus it is a many-fermion quantity. However, the numerical
calculations show that the two-fermion continuum dominates the
low-temperature behaviour of the $xx$ dynamic structure factor.
Although $S_{xx}(\kappa,\omega)$ is not restricted to the
two-fermion continuum region in the $\kappa$-$\omega$ plane and
has nonzero value above the upper boundary of the two-fermion
continuum (demonstrating the effects of the many-fermion continua
\cite{011}) its value outside the two-fermion continuum is rather
small. $S_{xx}(\kappa,\omega)$ may be described by several
washed-out excitation branches following roughly the two-fermion
continuum boundaries. Further studies are required to clarify why
two-fermion features rule the many-fermion quantity
$S_{xx}(\kappa,\omega)$. For $\vert\Omega\vert>\vert J\vert$ the
zero-temperature $xx$ dynamic structure factor
$S_{xx}(\kappa,\omega)$ shows a single $\delta$-peak along the
fermion branch $\omega=\Lambda_\kappa$. In the high-temperature
limit the low-temperature structures in $\kappa$-$\omega$ plane
disappear and $S_{xx}(\kappa,\omega)$ becomes $\kappa$-independent
in agreement with exact calculations for $\beta=0$ \cite{012}.
Alternatively the $xx$ dynamics can be examined using a
bosonization treatment \cite{013}, however, such an analysis is
restricted to low-energy physics and only a small region in the
$\kappa$-$\omega$ plane can be explored by this approach.

The described analysis of the dynamic properties may be extended
to the ani\-so\-tro\-pic $XY$ interaction ($J^x\ne J^y$)
\cite{008} and the dimerized isotropic $XY$ interaction
($J^x=J^y\to J(1-(-1)^n\delta)$, $0\leqslant\delta\leqslant 1$
is the
dimerization parameter) \cite{008,009}. Apparently, the exhaustive
study of the $zz$ dynamics for the former case \cite{008} should
be still supplemented by the corresponding analysis of the $xx$
dynamics whereas the case when both anisotropy and dimerization
are present requires a separate study. Recently, the effects of
periodic inhomogeneity on the dynamic susceptibility
$\chi_{zz}(\kappa,\omega)$ (but not $\chi_{xx}(\kappa,\omega)$)
have been reported in the paper on dynamics of isotropic $XY$
model on one-dimensional superlattices \cite{014}. The $xx$
dynamic quantities for the case of extremely anisotropic $XY$
interaction ($J^y=0$), i.e., for the spin-$\frac{1}{2}$ transverse
Ising chain, are of interest for interpreting the experimental
data on the dynamic dielectric permittivity of the
quasi-one-dimensional hydrogen-bonded ferroelectric compound
CsH$_2$PO$_4$ \cite{015,016}.

To end up this section, let us note that the dynamic properties of
two-dimensional quantum spin models can be also explained in terms
of the two-fermion continuum, however, such a picture may have
only an approximate meaning \cite{017}. Another interesting
question is to contrast the results for dynamic structure factors
of spin-$\frac{1}{2}$ and spin-$1$ chains \cite{018}.

\section{Dzyaloshinskii-Moriya interaction}

The Dzyaloshinskii-Moriya interaction is often present
in the low-dimensional quantum magnets
(see, for example, a recent paper \cite{019}).
It is generally known that the Dzyaloshinskii-Moriya interaction
$\sum_n{\bf {D}}\cdot ({\bf{s}}_n\times {\bf{s}}_{n+1})$
being added to the Hamiltonian (\ref{01})
does not destroy the rigorous treatment
if ${\bf {D}}=(0,0,D)$
\cite{020}\footnote
{Let us note
that in some cases
the Dzyaloshinskii-Moriya interaction
can be eliminated by the corresponding spin-coordinate transformation
\cite{021}
(see, for example, equation~(\ref{04})).}.

The main effect of the Dzyaloshinskii-Moriya interaction
is the loss of the symmetry of elementary excitation energies
with respect to the change $\kappa\to -\kappa$:
\begin{eqnarray}
\label{03}
\Lambda_{\kappa}
=D\sin\kappa
+\sqrt{\left(\Omega+\frac{J^x+J^y}{2}\cos\kappa\right)^2
+\left(\frac{J^x-J^y}{2}\right)^2\sin^2\kappa}
\ne \Lambda_{-\kappa}
\nonumber
\end{eqnarray}
\cite{020,022}.
In the presence of the Dzyaloshinskii-Moriya interaction
some remarkable changes
in the thermodynamic and dynamic properties of spin chains
may occur.
Consider, for example, the isotropic $XY$ chain.
Since the Dzyaloshinskii-Moriya interaction for such a chain
(even inhomogeneous one)
can be eliminated
by the local rotations in spin space around the $z$ axis
\begin{eqnarray}
\label{04}
s_n^x\cos\phi_n + s_n^y\sin\phi_n \to s_n^x,
\;\;\;
-s_n^x\sin\phi_n + s_n^y\cos\phi_n \to s_n^y\,,
\nonumber\\
\phi_n=\varphi_1+\ldots +\varphi_{n-1},
\;\;\;
\tan\varphi_m=\frac{D_m}{J_m}
\end{eqnarray}
resulting in a model with isotropic $XY$ interaction $\sqrt{J_n^2+D_n^2}$
the $zz$ dynamics remains as for a chain
without Dzyaloshinskii-Moriya interaction,
however,
with renormalized energy scale $J\to\sqrt{J^2+D^2}$~%
\footnote{This is not the case if $XY$ interaction is anisotropic;
the $zz$ dynamics of the model with extremely anisotropic $XY$ interaction,
i.e., the transverse Ising chain
with Dzyaloshinskii-Moriya interaction,
was considered in \cite{022}.}%
.
In contrast,
the $xx$ dynamic quantities according to (\ref{04}) involve
\begin{eqnarray}
\label{05}
\langle s_n^x(t)s_{n+m}^x \rangle_{J, D}
=& &
\cos\phi_n\cos\phi_{n+m}
\langle s_n^x(t)s_{n+m}^x \rangle_{\sqrt{J^2+D^2}, 0}
\nonumber\\
&-&
\cos\phi_n\sin\phi_{n+m}
\langle s_n^x(t)s_{n+m}^y \rangle_{\sqrt{J^2+D^2}, 0}
\nonumber\\
&-&
\sin\phi_n\cos\phi_{n+m}
\langle s_n^y(t)s_{n+m}^x \rangle_{\sqrt{J^2+D^2}, 0}
\nonumber\\
&+&
\sin\phi_n\sin\phi_{n+m}
\langle s_n^y(t)s_{n+m}^y \rangle_{\sqrt{J^2+D^2}, 0}
\end{eqnarray}
(for the homogeneous chain $\phi_n=(n-1)\varphi$,
$\tan\varphi=D/J$). In view of (\ref{05}) the relation
between the $zz$ and $xx$ dynamics discussed in the previous
section ($S_{xx}(\kappa,\omega)$ at low temperatures exhibits
the washed-out excitation branches which roughly follow the
boundaries of the two-fermion continuum which determines
$S_{zz}(\kappa,\omega)$) may appear to be more intricate
\cite{023}. In the next sections~4 and 5 we give further examples
of how the Dzyaloshinskii-Moriya interaction manifests itself in
the properties of spin-$\frac{1}{2}$ $XY$ chains.

\section{Correlated off-diagonal and diagonal disorder}

The Jordan-Wigner transformation maps
the spin-$\frac{1}{2}$ isotropic $XY$ chain in a transverse field
onto
the chain of tight-binding spinless fermions
with the on-site energy $\Omega$ and hopping $I=J/2$.
If the transverse fields are independent random variables
(diagonal disorder)
each with the Lorentzian probability distribution
\[
p(\Omega_n)=\frac{1}{\pi}\frac{\Gamma}{\left(\Omega_n-\Omega_0\right)^2+\Gamma^2}
\]
the resulting fermionic model
is the one-dimensional version of the Lloyd model.
The density of states,
\begin{eqnarray}
\overline{\rho(E)}
=\overline{\frac{1}{N}\sum_{k=1}^N\delta(E-\Lambda_k)},
\;\;\;
\overline{(\ldots)}
=\ldots\int{\mbox{d}}\Omega_np(\Omega_n)
\ldots(\ldots),
\nonumber
\end{eqnarray}
for the Lloyd model can be found exactly \cite{024}. Going far
beyond the idea of H.Ni\-shi\-mo\-ri we may consider a spin model with
the correlated off-diagonal and diagonal Lorentzian disorder which
after fermionization reduces to the one-dimensional version of the
extended Lloyd model introduced by W.John and J.Schreiber. Namely,
we consider the isotropic $XY$ model with independent random
exchange interactions (off-diagonal disorder) given by the
Lorentzian probability distribution
\begin{eqnarray}
\label{06}
p(\ldots,J_n,\ldots)
=\prod_np(J_n)
=\prod_n\frac{1}{\pi}\frac{\Gamma}{\left(J_n-J_0\right)^2+\Gamma^2}.
\end{eqnarray}
Moreover,
we consider the correlated off-diagonal and diagonal disorder
assuming that
the on-site transverse fields in the chain
are determined by the surrounding exchange interactions
according to the relation
\begin{eqnarray}
\label{07}
\Omega_n-\Omega_0
=\frac{a}{2}\left(J_{n-1}+J_n-2J_0\right),
\;\;\;
a\;{\mbox{is real}},
\;\;\;
\vert a\vert\geqslant 1.
\end{eqnarray}
Then the density of states $\overline{\rho(E)}$
yielding the thermodynamic quantities
of the introduced random spin chain
can be calculated exactly
\cite{025,026}.
To get $\overline{\rho(E)}$
we must calculate the diagonal Green functions
$\overline{G_{nn}^{\mp}(E)}$,
since
\[
\overline{\rho(E)}
=\mp\frac{1}{\pi}\Im\overline{G_{nn}^{\mp}(E)}
.
\]
The set of equations of motion for
$G_{nm}^{\mp}(E\pm{\mbox{i}}\epsilon)$, $\epsilon\to+0$
can be averaged using contour integration
in complex planes of random (Lorentzian) variables $J_n$.
Using the Gershgorin criterion
we find the set of equations for the averaged Green functions
which has the same structure as before averaging
but possesses translational symmetry.
As a result we obtain the desired quantities
$\overline{G_{nm}^{\mp}(E)}$
and hence all thermodynamic quantities.

\begin{figure}[h]
\centerline{\includegraphics[width=0.78\textwidth]{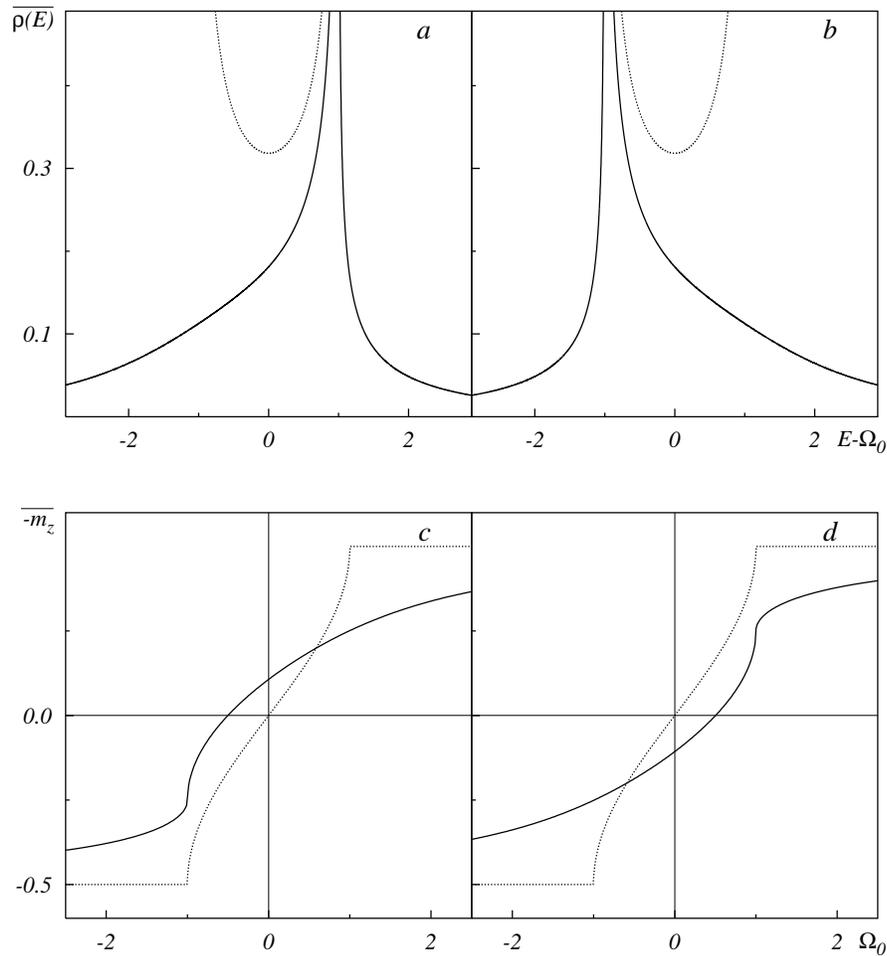}}
\caption{The density of states (a, b)
and the transverse magnetization (c, d)
of the isotropic $XY$ chain with correlated Lorentzian disorder
($J_0=1$, $\Gamma=1$,
$a=-1$ (a, c),
$a=1$ (b, d)).
The dotted curves correspond to the nonrandom case ($\Gamma=0$).}
\label{fig02}
\end{figure}

In figures~2a and 2b
we present $\overline{\rho(E)}$ in the most interesting region
$\vert a\vert\to 1$, when $\overline{\rho(E)}$ becomes not
symmetric with respect to the change $E-\Omega_0\to
-\left(E-\Omega_0\right)$. Such asymmetry immediately yields a
nonzero (average) transverse magnetization
\[
\overline{m_z}
=-\frac{1}{2}\int {\mbox{d}}E\overline{\rho(E)}\tanh\frac{\beta
E}{2}\ne 0
\]
at low temperatures $\beta\to \infty$ at zero
(average) transverse field $\Omega_0=0$ (figures~2c and 2d). Let us
consider more closely this somewhat unexpected magnetic property
of the introduced random spin chain. For a certain random
realization of the chain defined by (\ref{06}), (\ref{07}) one may
expect the same numbers of sites surrounded by stronger than $J_0$
exchange interactions as the sites surrounded by weaker than $J_0$
exchange interactions. Because of (\ref{07}) for $\Omega_0=0$ the
transverse fields at the former and at the latter sites have the
same value but the opposite signs giving as a result
$\sum_n\Omega_n=0$. On the other hand, one may expect that the
sites surrounded by strong isotropic $XY$ exchange interactions
exhibit small $z$ magnetization whereas the sites surrounded by
weak isotropic $XY$ exchange interactions exhibit large $z$
magnetization (in the opposite direction). As a result, the
average transverse magnetization has a nonzero value. As $\vert
a\vert$ increases, a difference in the oppositely directed $z$
magnetizations becomes smaller. Thus, a nonzero $\overline{m_z}$
at $\Omega_0=0$ appears owing to the imposed relation (\ref{07})
which expresses the condition of correlated disorder.

Some further insight into the origin of the asymmetry of
$\overline{\rho(E)}$ can be obtained after examining the moments
of the density of states
\begin{eqnarray}
\label{08}
M^{(r)}
&\equiv&\int{\mbox{d}}EE^r\rho(E)
\nonumber\\
&=&\frac{1}{N}
\sum_{n=1}^N
\left\langle
\left\{
\left[
\ldots
\left[c_n,H\right],
\ldots
,
H
\right],
c_n^+
\right\}
\right\rangle
\end{eqnarray}
(here $H$ is the Hamiltonian of fermions
which represent the spin chain).
It is just the correlated disorder that yields a nonzero third moment
$\overline{M^{(3)}}\ne 0$\footnote
{Here $\overline{(\ldots)}$
denotes the random averaging
either
for the correlated off-diagonal and diagonal disorder
or
for the independent off-diagonal
(with the probability distribution $p(J_n)$)
and diagonal
(with the probability distribution
$p(\Omega_n)=\int{\mbox{d}}J_{n-1}\int{\mbox{d}}J_n
p(J_{n-1})p(J_n)
\delta\left(\Omega_n-\Omega_0-\frac{a}{2}
\left(J_{n-1}+J_n
\right.\right.
$
$
\left.\left.
-2J_0\right)\right)$) disorders.}
at $\Omega_0=0$. For not correlated off-diagonal and diagonal
disorders one gets $\overline{M^{(3)}}=0$ at $\Omega_0=0$. The
moments of the density of states can be calculated for any
probability distribution of random variables $p(J_n)$ (not
necessarily for the Lorentzian probability distribution), for
example, for the rectangle probability distribution. These
calculations explicitly demonstrate the cause of the asymmetry
appearance in $\overline{\rho(E)}$ \cite{027}. Some other results
on the effects of correlated disorder can be found in
\cite{010,028,029}.

Finally, let us remark that the considered spin model with
correlated Lorentzian disorder may be extended by introducing the
nonrandom Dzyaloshinskii-Moriya interaction $D$. Another extension
is to assume the exchange interaction to be nonrandom $J_n=J$
whereas the Dzyaloshinskii-Moriya interactions $D_n$ to be
independent random Lorentzian variables determining the transverse
fields according to (\ref{07}). Both models are related to each
other through a certain sequence of rotations of spin axes around
the $z$ axis (being applied to the Hamiltonian with the exchange
interactions $D_n$ and the Dzyaloshinskii-Moriya interactions
$-J_n$ it gives the Hamiltonian with the exchange interactions
$J_n$ and the Dzyaloshinskii-Moriya interactions $D_n$) and hence
it is sufficient to consider only one of them. Considering, for
example, the former model one finds that the nonrandom
Dzyaloshinskii-Moriya interaction may lead to the recovery of the
symmetry with respect to the change $E-\Omega_0\to
-\left(E-\Omega_0\right)$ and hence to a decrease of the nonzero
value of $\overline{m_z}$ at $\Omega_0=0$. Such an effect becomes
also apparent after calculating the moments of the density of
states $\overline{M^{(2)}}$ and $\overline{M^{(3)}}$ (\ref{08})
for this model.

To end up this section, let us note that a different random
spin-$\frac{1}{2}$ $XY$ chains were rigorously analytically
examined by Th.M.Nieuwenhuizen and coauthors \cite{030}.

\section{Effects of regularly alternating bonds and fields}

\subsection{Continued fractions}

The quantum spin chains with regularly alternating Hamiltonian parameters
can model dimerized ($l$-merized) chains,
one-dimensional superlattices,
one-dimensional decorated chains etc.
The case of spin-$\frac{1}{2}$ $XY$ chains
supplemented by the continued fraction approach
(for other approaches see \cite{031})
remains amenable for rigorous analysis
of the thermodynamic properties
if the exchange interaction is isotropic
($J_n^x=J_n^y$) \cite{032},
extremely anisotropic
($J_n^y=0$) \cite{033}
or if $\Omega_n=0$ \cite{034}.
The thermodynamic quantities
of the regularly alternating isotropic $XY$ chain in a transverse field
can be obtained through the density of states
$\rho (E)=\frac{1}{N}\sum_{k=1}^N\delta(E-\Lambda_k)$.
The thermodynamic quantities
of the regularly alternating transverse Ising chain
and the regularly alternating
anisotropic $XY$ chain without transverse field
can be obtained through the density of states
$R(E^2)=\frac{1}{N}\sum_{k=1}^N\delta(E^2-\Lambda_k^2)$.

Let us recall
that after fermionization
of the isotropic $XY$ chain in a transverse field
($J_n^x=J_n^y=2I_n$)
one faces the Hamiltonian
which is a bilinear fermion form.
While making it diagonal
one arrives at the set of equations
\begin{eqnarray}
\label{09} I_{n-1}g_{k,n-1} +\left(\Omega_n-\Lambda_k\right)g_{kn}
+I_ng_{k,n+1} =0,
\end{eqnarray}
here $g_{kn}$ are the coefficients of the linear transformation
which make the initial bilinear fermion form diagonal
and
$\Lambda_k$ are the resulting elementary excitation energies.
Therefore,
introducing the Green functions ${\cal{G}}_{nm}(E)$
according to
\begin{eqnarray}
\label{10}
-I_{n-1}{\cal{G}}_{n-1,m}(E)
+\left(E-\Omega_n\right){\cal{G}}_{nm}(E)
-I_n{\cal{G}}_{n+1,m}(E)
=\delta_{nm}
\end{eqnarray}
one gets the density of states
\[
\rho(E)=\mp\frac{1}{\pi N}\sum_{n=1}^N\Im{\cal{G}}_{nn}(E\pm{\mbox{i}}\epsilon)
,
\quad
\epsilon\to +0
.
\]
For the transverse Ising chain
($J_n^x=2I_n$, $J_n^y=0$)
instead of (\ref{09}) and (\ref{10})
we have
\begin{eqnarray}
\label{11}
\Omega_{n-1}I_{n-1}\Phi_{k,n-1}
+\left(\Omega_n^2+I_{n-1}^2-\Lambda_k^2\right)\Phi_{kn}
+\Omega_nI_n\Phi_{k,n+1}
=0
\end{eqnarray}
and
\begin{eqnarray}
\label{12} -\Omega_{n-1}I_{n-1}{\cal{G}}_{n-1,m}(E^2)
+\left(E^2-I_{n-1}^2-\Omega_n^2\right){\cal{G}}_{nm}(E^2)
-\Omega_nI_n{\cal{G}}_{n+1,m}(E^2) =\delta_{nm}\,,
\end{eqnarray}
respectively,
whereas for the anisotropic $XY$ chain without field
($J_n^x=2I_n^x$,
$J_n^y=2I_n^y$,
$\Omega_n=0$)
instead of (\ref{09}) and (\ref{10})
we have
\begin{eqnarray}
\label{13}
I_{n-2}^yI_{n-1}^x\Phi_{k,n-2}
+\left({I_{n-1}^x}^2+{I_n^y}^2-\Lambda_k^2\right)\Phi_{kn}
+I_n^yI_{n+1}^x\Phi_{k,n+2}
=0
\end{eqnarray}
and
\begin{equation}
\label{14}
-I_{n-2}^yI_{n-1}^x{\cal{G}}_{n-2,m}(E^2)
+\left(E^2-{I_{n-1}^x}^2-{I_n^y}^2\right){\cal{G}}_{nm}(E^2)
-I_n^yI_{n+1}^x{\cal{G}}_{n+2,m}(E^2)
=\delta_{nm}\,,
\end{equation}
respectively.
For the last two models the Green functions
${\cal{G}}_{nm}(E^2)$
yield
the density of states
\[
R(E^2)
=\mp\frac{1}{\pi N}\sum_{n=1}^N\Im{\cal{G}}_{nn}(E^2\pm{\mbox{i}}\epsilon)
,\quad
\epsilon\to +0
.
\]
For a general case of the anisotropic $XY$ chain
in a transverse field
a set of equations like (\ref{11}) or (\ref{13})
is five diagonal banded
(but not three diagonal banded as (\ref{09}), (\ref{11}) or (\ref{13}))
and the next step,
i.e., the continued fraction representation for the diagonal Green functions
is less evident.
According to equation~(\ref{10}), (\ref{12}) or (\ref{14})
the diagonal Green functions for all these models
can be represented in terms of continued fractions.
For example,
from (\ref{14}) it immediately follows that
\begin{eqnarray}
{\cal{G}}_{nn}(E^2)
=\frac{1}{E^2-{I_{n-1}^x}^2-{I_n^y}^2-\Delta_n^--\Delta_n^+},
\nonumber\\
\Delta_n^-
=\frac{{I_{n-2}^y}^2{I_{n-1}^x}^2}
{E^2-{I_{n-3}^x}^2-{I_{n-2}^y}^2-
\frac{{I_{n-4}^y}^2{I_{n-3}^x}^2}{E^2-{I_{n-5}^x}^2-{I_{n-4}^y}^2-_{\ddots}}},
\nonumber\\
\Delta_n^+
=\frac{{I_{n}^y}^2{I_{n+1}^x}^2}
{E^2-{I_{n+1}^x}^2-{I_{n+2}^y}^2-
\frac{{I_{n+2}^y}^2{I_{n+3}^x}^2}{E^2-{I_{n+3}^x}^2-{I_{n+4}^y}^2-_{\ddots}}}.
\label{15}
\end{eqnarray}
If now the Hamiltonian parameters are periodic with any finite
period $p$ the continued fractions $\Delta_n^{\mp}$ in (\ref{15})
become periodic and can be evaluated by solving square equations.
As a result, we get the exact expressions for
${\cal{G}}_{nn}(E^2)$ (or ${\cal{G}}_{nn}(E)$ for the isotropic
$XY$ chain) and hence for all thermodynamic quantities of the
regularly alternating spin chains in question.

Let us remark, that the thermodynamic quantities of the
anisotropic $XY$ chain in a transverse field of period 2 were
obtained in \cite{035} in a different manner on a quite general
background. This scheme, however, becomes cumbersome if the period
of nonuniformity increases. Further development of a general
approach to the study of thermodynamic quantities and spin
correlation functions has been reported by L.L.Gon\c{c}alves with
coworkers in connection with spin-$\frac{1}{2}$ $XY$ models on
one-dimensional superlattices.

\subsection{Magnetization processes}

The most spectacular manifestation of the effects of regular
alternation can be seen in the magnetization processes at low
temperatures. Consider at first the isotropic $XY$ chain. Owing to
regularly alternating parameters of the Hamiltonian the fermion
band  splits into several subbands the number of which does not
exceed the period of regular inhomogeneity. This circumstance
immediately suggests that the zero-temperature dependence
\[
m_z=-\frac{1}{2}\int{\mbox{d}}E\rho(E)\tanh\frac{\beta E}{2}
\]
on
$\Omega$ (we have assumed that $\Omega_n=\Omega+\Delta\Omega_n$)
consists of horizontal parts (plateaus) which appear when $E=0$
remains in between subbands (given by $\rho(E)$) with changing of
$\Omega$ separated by the parts of varying $m_z$ which appear when
$E=0$ remains within subbands with $\Omega$ changing. Clearly, the
number of plateaus does not exceed the period of nonuniformity.
Moreover, their heights $-s\leqslant m_z\leqslant s$ are in agreement with the
condition: $p\left(s-m_z\right)={\mbox{integer}}$ with
$s=\frac{1}{2}$ \cite{036}. This conjecture was suggested on the
model-independent background for a general spin-$s$ chain with
axial symmetry in a uniform magnetic field using the
Lieb-Schultz-Mattis theorem and the bosonization arguments.
Obviously, spin-$\frac{1}{2}$ $XY$ chain is an easy case
(represented by noninteracting fermions) contrary to the chains
with Heisenberg exchange interaction or higher spins
$s>\frac{1}{2}$. On the other hand, the magnetization curve for
spin-$\frac{1}{2}$ $XY$ chain can be obtained explicitly.
Moreover, the elaborated scheme also permits to obtain the local
(on-site) magnetizations. The on-site magnetizations exhibit
plateaus which begin and end up at the same values of $\Omega$ as
for the total magnetization, however, the plateaus heights (i.e.,
the values of the on-site magnetizations) are not universal
quantities no longer obeying the above-mentioned condition
and strongly depend on a concrete set of the Hamiltonian
parameters. Moreover, a sequence of sites $n_1,n_2,\ldots,n_p$
satisfying the inequalities
${m_z}_{n_1} \leqslant {m_z}_{n_2} \leqslant\ldots
\leqslant {m_z}_{n_p}$
depends on the value of the applied field
$\Omega$. The step-like behaviour of the zero-temperature
dependence $m_z$ (${m_z}_n$) vs. $\Omega$ is accompanied by the
corresponding zero-temperature behaviour
$\chi_z={\partial m_z}/{\partial \Omega}$
(${\chi_z}_n={\partial {m_z}_n}/{\partial \Omega}$) vs.
$\Omega$. The zero-temperature
dependence $\chi_z$ on $\Omega$ behaves like $-\rho(E=0)$ vs.
$\Omega$ thus reproducing the density of states $\rho(E)$.

In contrast to the isotropic $XY$ chain, a regular alternation of
the parameters of the transverse Ising chain  Hamiltonian does not
lead to plateaus in the zero-temperature dependence $m_z$ on
$\Omega$. The difference arises owing to the anisotropy of $XY$
interaction: if $J_n^x\ne J_n^y$ one finds that $\left[\sum_n
s_n^z, H\right]\ne 0$. Thus, even if the spin system remains in
the same ground state $\vert 0\rangle$ with varying $\Omega$ the
magnetization $\langle 0\vert \sum_ns_n^z\vert 0\rangle$,
nevertheless, varies with varying $\Omega$. As a result the
dependence $m_z$ vs. $\Omega$ does not exhibit horizontal parts.
On the other hand, in the vicinity of some fields $\Omega^{\star}$
(critical fields) the magnetization may behave as $m_z\sim
\left(\Omega-\Omega^{\star}\right)
\ln\vert\Omega-\Omega^{\star}\vert$ and hence the susceptibility
as $\chi_z\sim\ln\vert\Omega-\Omega^{\star}\vert$. For the uniform
chain there are two critical fields $\Omega^{\star}=\pm \vert
I\vert$. If the bonds become regularly alternating only the values
of $\Omega^{\star}$ change. For example,
$\Omega^{\star}=\pm\sqrt{\vert I_1I_2\vert}$ for $p=2$,
$\Omega^{\star}=\pm\sqrt[3]{\vert I_1I_2I_3\vert}$ for $p=3$ etc.
If the fields become regularly alternating not only the values of
the critical fields vary but the number of the critical fields may
change. For example, assuming
$\Omega_{1,2}=\Omega\pm\Delta\Omega$, $\Delta\Omega\geqslant 0$ one has
either two critical fields
$\Omega^{\star}=\pm\sqrt{{\Delta\Omega}^2+\vert I_1I_2\vert}$ if
$\Delta\Omega<\sqrt{\vert I_1I_2\vert}$ or four critical fields
$\Omega^{\star} =\pm\sqrt{{\Delta\Omega}^2\pm\vert I_1I_2\vert}$
if $\Delta\Omega>\sqrt{\vert I_1I_2\vert}$. The transverse field
is a controlling parameter of the second-order quantum phase
transition in the spin-$\frac{1}{2}$ Ising chain in a transverse
field \cite{037}. Thus, a number of the quantum phase transition
points governed by $\Omega$  in this model may increase due to a
regular alternation of the transverse fields presuming the
strength of inhomogeneity is sufficiently strong. For example, for
$p=2$ there may be either two or four quantum phase transition
points, for $p=3$ there may be either two, or four, or six quantum
phase transition points etc. The critical behaviour remains
unchanged and is just like for the temperature driven phase
transition in the square-lattice Ising model. For example, the
zero-temperature dependence $\chi_z$ vs. $\Omega$ always exhibits
logarithmic singularities the number of which depends on a
concrete set of the Hamiltonian parameters. It seems interesting
to trace the changes in magnetization processes as anisotropy of
the exchange interaction varies between the isotropic and
extremely anisotropic limits.

\begin{wrapfigure}[35]{i}{0.4\textwidth}
\vspace{-3ex}
\centerline{\includegraphics[width=0.36\textwidth]{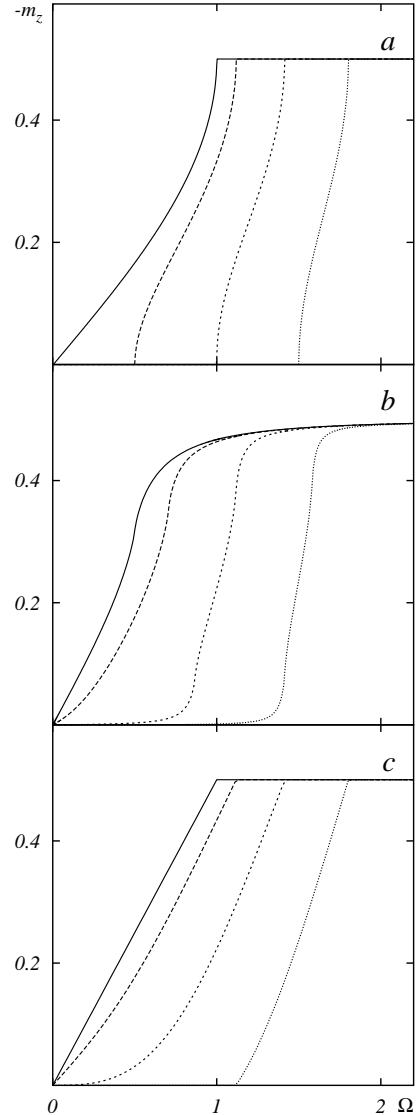}}
\caption{%
The zero-temperature magnetization curves
for the isotropic $XY$ (a),
transverse Ising (b)
and classical (c) chains of period 2
($J_n=1$,
$\Omega_{1,2}=\Omega\pm\Delta\Omega$,
$\Delta\Omega=0$ (solid curves),
$\Delta\Omega=0.5$ (long-dashed curves),
$\Delta\Omega=1$ (short-dashed curves),
$\Delta\Omega=1.5$ (dotted curves)).}%
\label{fig03}%
\end{wrapfigure}
It is instructive to compare
the zero-temperature magnetization processes
in the quantum and classical $XY$ chains\footnote
{Rigorous calculations
of the initial static susceptibility tensor
for another nonuniform classical chain,
i.e., the spin-$\frac{1}{2}$ Ising chain,
can be found in \cite{038}.}.
A classical chain
consists of arrows (vectors)
${\bf {s}}=(s,\theta,\phi)$,
$s=\frac{1}{2}$,
$0\leqslant\theta\leqslant\pi$,
$0\leqslant\phi<2\pi$
which interact with each other and an external (transverse) field.
The classical isotropic $XY$ chain is described by the Hamiltonian
\begin{eqnarray}
\lefteqn{H=\sum_n\Omega s\cos\theta_n}
&&\nonumber\\
&&\!\!\!{}
+\sum_n J s^2\sin\theta_n\sin\theta_{n+1}\cos\left(\phi_n-\phi_{n+1}\right)
\qquad\mathstrut
\label{16}
\end{eqnarray}
(compare with (\ref{01})).
The Hamiltonian of the classical transverse Ising chain
contains
$\cos\phi_n\cos\phi_{n+1}$
instead of
$\cos\left(\phi_n-\phi_{n+1}\right)$
in the interspin interaction terms in (\ref{16}).
Consider further a chain of period 2
with
$\Omega_{1,2}=\Omega\pm\Delta\Omega$, $\Delta\Omega\geqslant 0$,
$J_{1,2}=J$.
The ground-state energy ansatz
for both the isotropic $XY$ and transverse Ising chains
reads
\begin{eqnarray}
\lefteqn{%
E_0(\theta_1,\theta_2)=\frac{N}{2}s
\left(\left(\Omega+\Delta\Omega\right)\cos\theta_1\right.}
&&\nonumber\\
&&\!\!\!{}
+\left.\left(\Omega-\Delta\Omega\right)\cos\theta_2\right)
-N\vert J\vert s^2\sin\theta_1\sin\theta_2
\qquad\mathstrut
\label{17}
\end{eqnarray}
and $\theta_1$, $\theta_2$ are determined from the equations
\begin{eqnarray}
\left(\Omega+\Delta\Omega\right)\sin\theta_1
+2\vert J\vert s\cos\theta_1\sin\theta_2&=&0,
\nonumber\\
\left(\Omega-\Delta\Omega\right)\sin\theta_2
+2\vert J\vert s\sin\theta_1\cos\theta_2&=&0
\qquad\mathstrut
\label{18}
\end{eqnarray}
to provide a minimum of $E_0(\theta_1,\theta_2)$ (\ref{17}).
The on-site magnetizations
have the component
arbitrary directed in the $xy$ plane
with $\vert {m_{\perp}}_n\vert=s\sin\theta_n$
($n=1,2$)
for the isotropic $XY$ case
or
the component directed
along the $x$ axis
with
$\vert {m_x}_n\vert=s\sin\theta_n$
for the transverse Ising case
and
the $z$ component given by
$m^z_n=s\cos\theta_n$
in both cases.
The components of the total magnetization per site
are then as follows
$m_{\perp}=\frac{1}{2}\left({m_\perp}_1+{m_\perp}_2\right)$
or
$m_x=\frac{1}{2}\left({m_x}_1+{m_x}_2\right)$
(if $J<0$)
and
$m_z=\frac{1}{2}\left({m_z}_1+{m_z}_2\right)$.
Equations~(\ref{18}) may be solved analytically
yielding the ground-state energy
and the on-site magnetizations
${m_\perp}_1$ (${m_x}_1$),
${m_\perp}_2$ (${m_x}_2$),
${m_z}_1$,
${m_z}_2$.
In figure~3
we contrast the zero-temperature magnetization curves
for the isotropic $XY$, transverse Ising
and classical chains
of period 2.
Note,
that the classical isotropic $XY$ chain
similarly to its quantum counterpart
may exhibit plateaus in the dependence $m_z$ vs. $\Omega$.
Evidently,
it would be interesting to examine a quantum-to-classical crossover
in the magnetization curves.
Probably a real-space renormalization-group
method
aimed on the study of quantum fields on a lattice
and applied
in \cite{039}
to the spin-$\frac{1}{2}$ transverse Ising chain
may be used to analyse the spin-$1$ chain,
that is,
an intermediate case between the quantum ($s=\frac{1}{2}$)
and the classical ($s\to\infty$) cases.

Exact results
for the magnetization curves of spin-$\frac{1}{2}$ $XY$ chains
may be employed
to test the strong-coupling approach
that is a widely used
approximate approach
in the theory of low-dimensional spin systems
consisting of
periodically repeating units
(spin chains with a periodic modulation
of the intersite interactions
or spin ladders)
\cite{040}.

\subsection{Spin-Peierls instability}

Since the paper of P.Pincus \cite{041} we know that the
spin-$\frac{1}{2}$ isotropic $XY$ chain is unstable with respect
to a lattice dimerization due to the spin-Peierls
mechanism\footnote {This conclusion was obtained within the
adiabatic approximation; the corresponding treatment within the
nonadiabatic approximation is more sophisticated \cite{042}.}.
Really, the ground-state energy per site of the dimerized spin
chain (i.e., with $J_n=J\left(1-(-1)^n\delta\right)$) is given by
\begin{eqnarray}
\label{19}
e_0(\delta)
=-\frac{\vert J\vert}{\pi}
\int_0^{\psi}
{\mbox{d}}\varphi\sqrt{1-\left(1-\delta^2\right)\sin^2\varphi}
-
\vert\Omega\vert\left(\frac{1}{2}-\frac{\psi}{\pi}\right),
\\
\psi=\left\{
\begin{array}{ll}
0, &
{\mbox{if}}\;\;\;\vert J\vert\leqslant\vert\Omega\vert,\\
\arcsin{\sqrt{\frac{1-{\Omega^2}/{J^2}}{1-\delta^2}}}, &
{\mbox{if}}\;\;\;\delta\vert J\vert\leqslant\vert\Omega\vert<\vert J\vert,\\
\frac{\pi}{2}, &
{\mbox{if}}\;\;\;\vert\Omega\vert<\delta\vert J\vert.
\end{array}
\right.
\nonumber
\end{eqnarray}
At $\Omega=0$ the ground-state energy per site (\ref{19})
($e_0(\delta) \approx e_0(0) +\frac{\vert
J\vert}{2\pi}\delta^2\ln\delta$, $\delta\ll 1$) decreases rapidly
enough in comparison with the increase of the elastic energy per
site $\alpha\delta^2$ as $\delta$ increases to provide a minimum
of the total energy per site $e_0(\delta)+\alpha\delta^2$ at a
nonzero value of the dimerization parameter $\delta^{\star}$. The
external field may destroy dimerization: if $\Omega$ exceeds a
certain value the dimerized phase cannot survive and the uniform
phase becomes favourable. The phase diagram shown in figure~4a
specifies the region of stability/metastability of the dimerized phase
at zero temperature.

It is generally known
(see, for example, a review on CuGeO$_3$ \cite{043})
that the increase of external field leads to a transition
from the dimerized phase to the incommensurate phase
rather than to the uniform phase
that contradicts to what is seen in figure~4.
Obviously, the incommensurate phase cannot appear within the frames
of the adopted ansatz for lattice distortion
$\delta_1\delta_2\delta_1\delta_2\ldots\;$,
$\delta_1+\delta_2=0$.
Therefore, we
\begin{wrapfigure}[44]{i}{0.46\textwidth}
\centerline{\includegraphics[width=0.38\textwidth]{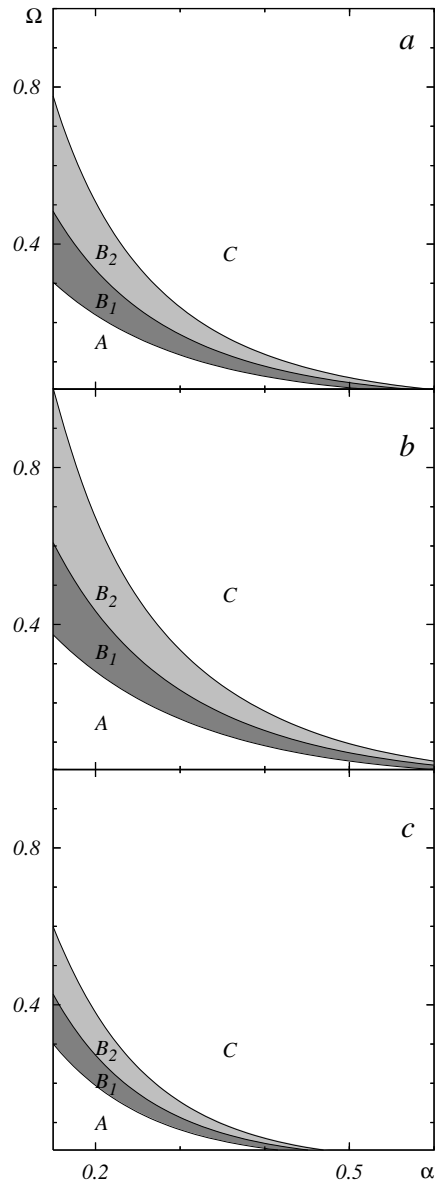}}
\caption{%
The phase diagram of the isotropic $XY$ chain ($J=1$)
in the plane
lattice stiffness $\alpha$ -- transverse field $\Omega$
which indicates
the regions of stability/metastability
of the dimerized and uniform phases
at zero temperature.
A (C) -- only dimerized (uniform) phase occurs;
B$_1$ (B$_2$) both phases are possible but the dimerized (uniform)
is the favourable one.
The phase diagram also illustrates
the possible effects of the Dzyaloshinskii-Moriya interaction
(a: $D=0$,
b: $D=0.5$, $k=1$,
c: $D=0.5$, $k=0$).}%
\label{fig04}%
\end{wrapfigure}
should assume $J_n=J\left(1+\delta_n\right)$
and examine the total energy
$E_0\left(\left\{\delta_n\right\}\right)
+\sum_n\alpha\delta_n^2$
for different lattice distortions
$\left\{\delta_n\right\}$.
We introduce a trial distortion of the form
\begin{eqnarray}
\label{20} \delta_n=-\delta\cos\left(\frac{2\pi}{p}n\right),
\end{eqnarray}
where $p$ is the period of modulation and analyse numerically
($N=1000$) the total energy to reveal which spin-Peierls phase is
realized in the presence of external field (for details see
\cite{044}). Comparing the behaviour of the total energy for $p=2$
and $p=1.9$, $p=2.1$ as $\Omega$ increases we conclude that a
long-period structure does arise if $\Omega$ exceeds a certain
value. The dimerized phase transforms into a long-period phase
rather than into the uniform phase while the field increases. The
important conclusion of further computations using (\ref{20}) is
that the dimerized phase persists up to a certain characteristic
field. Further, in moderate fields the lattice parameterized by
(\ref{20}) may exhibit short-period phases, for example, the
trimerized phase with $p=3$ \cite{045} (another possible lattice
distortion which preserves the chain length is
$\delta_1=-\delta_2$). However, a behaviour of the trimerized
phase is essentially different in comparison with that of the
dimerized phase: for any small deviation of the field from the
value at which the trimerized phase occurs there exists such a
long-periodic structure for which the lattice distortion
(\ref{20}) gives smaller total energy than for $p=3$. Thus,
contrary to the dimerized phase, the trimerized phase does not
persist with the field varying and it continuously transforms into
a certain long-period phase with the field varying. In strong
fields the uniform lattice can be expected. Clearly, such a study
is restricted to the adopted ansatz (\ref{20}) (although any other
(but not all) distortion pattern may be assumed) and therefore we
can say for sure which lattice distortion is not realized rather
than to point out which lattice distortion should occur.

It is interesting to discuss the effect of the
Dzyaloshinskii-Moriya interaction on the spin-Peierls
dimerization. For this purpose we should find the ground-state
energy of the chain of period 2 with a sequence of parameters
$$
\Omega J_1D_1\Omega J_2D_2\Omega J_1D_1 \Omega J_2D_2\ldots\, .
$$
Moreover, $J_{1,2}=J(1\pm\delta)$ and
$D_{1,2}=D(1\pm k\delta)$. Putting $k=0$ one has a chain in which
$D$ does not depend on the lattice distortion, whereas for $k=1$
the dependence of $D$ on the lattice distortion is the same as
that for the isotropic $XY$ exchange interaction. In the limit of
small $\delta\ll 1$ (valid for lattices with large values of
$\alpha$ and corresponding  to the experimental situation) the
analysis becomes extremely simple. It is convenient to introduce
the parameter
\begin{eqnarray}
\nonumber
\aleph=
\frac{J^2+kD^2}{J^2+D^2}.
\end{eqnarray}
Consider, for example, the case $\Omega=0$.
After a simple rescaling of variables
one arrives at
the equations considered in \cite{041} and as a result
\begin{eqnarray}
\nonumber
\delta^{\star}
\sim\frac{1}{\aleph}\exp\left(-\frac{2\pi\alpha}
{\sqrt{J^2+D^2}\aleph^2}\right).
\end{eqnarray}
We immediately conclude that for $k=1$ nonzero $D$ leads to the
increase of the dimerization parameter $\delta^{\star}$, whereas
for $k=0$ nonzero $D$ leads to the decrease of $\delta^{\star}$.
Thus, the Dzyaloshinskii-Moriya interaction may act either in
favour of the dimerization or against it. The actual result of its
effect depends on the dependence of the Dzyaloshinskii-Moriya
interaction on the amplitude of the lattice distortion in
comparison with the corresponding dependence of the isotropic $XY$
exchange interaction. Comparing figure~4a and figures~4b, 4c one can
see the described effects of the Dzyaloshinskii-Moriya interaction
on the spin-Peierls dimerization.

Finally, let us make some remarks on the exchange interaction
anisotropy effects on the spin-Peierls dimerization. Consider the
anisotropic $XY$ chain (without field). In the isotropic limit it
exhibits the spin-Peierls dimerized phase, whereas in the
extremely anisotropic limit (Ising chain) the ground-state energy
does not depend on $\delta$ and the spin-Peierls dimerization
cannot occur. Therefore, it is attracting to follow how anisotropy
of the exchange interaction destroys the dimerized phase and to
find, in particular, a critical value of anisotropy above which
the dimerized phase cannot appear. The consideration becomes even
more complicated in the presence of an external (transverse)
field. Note, however, that such a study could be performed many
years ago on the basis of the results for the anisotropic $XY$
chain in a transverse field of period 2 reported in \cite{035}.

Another issue which deserves to be studied in detail is related to
a spin-Peierls instability of the spin-$\frac{1}{2}$ transverse
Ising chain \cite{033}. By direct inspection one can make sure
that although the ground-state energy of the dimerized Ising chain
in a constant transverse field $\Omega$ decreases as $\delta$
increases, however, not sufficiently rapidly (for any $\Omega$) to
get a gain in the total energy. As a result no spin-Peierls
dimerization should be anticipated. On the other hand, the
transverse Ising chain is unitary equivalent to the anisotropic
$XY$ chain without field. By comparison of the diagonal Green
functions (see equation~(\ref{15}) in the present paper
and equation~(8) in
\cite[J.~Phys.~A]{033}) it can be found
that the Helmholtz free energy of the anisotropic $XY$ chain
without field of $N$ sites $I^x_nI^y_nI^x_{n+1}I^y_{n+1}\ldots$ is
the sum of the Helmholtz free energies of two transverse Ising
chains $\Omega_nI_n\Omega_{n+1}I_{n+1}\ldots$ of ${N}/{2}$
sites, namely,
\[
\ldots I_{n-1}^xI_n^yI_{n+1}^xI_{n+2}^y\ldots
\]
and
\[
\ldots I_{n-1}^yI_n^xI_{n+1}^yI_{n+2}^x\ldots\,.
\]
Consider at first the following isotropic $XY$ chain
$I_n^x=I_n^y=I\left(1-(-1)^n\delta\right)$ which exhibits the
spin-Peierls dimerization. In view of the mentioned correspondence
the introduced model is thermodynamically equivalent to two
uniform transverse Ising chains (both with the transverse field
$I(1+\delta)$ (or $I(1-\delta)$) and the Ising exchange
interaction $I(1-\delta)$ (or $I(1+\delta)$)) and thus we conclude
that the critical Ising chain ($\Omega=I$) is unstable with
respect to a uniform extending (or shortening) accompanied by the
corresponding increase (decrease) of the transverse field.
Further, it is known that the quadrimerized isotropic $XY$ chain
without field may be energetically favourable in comparison with
the uniform chain (although yielding a smaller gain in the total
energy than the dimerized chain). Consider, therefore, such a
chain of period 4 with $I_n=I^{\prime}(1+\delta)$,
$I_{n+1}=I^{\prime\prime}(1+\delta)$,
$I_{n+2}=I^{\prime}(1-\delta)$,
$I_{n+3}=I^{\prime\prime}(1-\delta)$ which is unstable with
respect to a lattice distortion characterizing by nonzero value of
$\delta$ for certain values of $I^\prime$, $I^{\prime\prime}$.
This model exhibits the same thermodynamics as two identical
transverse Ising chains of period 2
$\Omega_nI_n\Omega_{n+1}I_{n+1}
\Omega_nI_n\Omega_{n+1}I_{n+1}\ldots$ with
$\Omega_n=I^{\prime\prime}(1-\delta)$,
$\Omega_{n+1}=I^{\prime\prime}(1+\delta)$ and
$I_n=I^{\prime}(1+\delta)$, $I_{n+1}=I^{\prime}(1-\delta)$. These
arguments indicate a possibility of the spin-Peierls bond
dimerization of the spin-$\frac{1}{2}$ transverse Ising chain
accompanied by the coherent modulation of the on-site transverse
fields.

\section{Skipped items and summary}

Needless to say that many important contributions have appeared
out of the scope of this brief survey. Let us discuss in a
telegraph-style manner some of them. Numerous works were devoted
to the analysis of the properties of spin-$\frac{1}{2}$ $XY$
chains with aperiodic or random Hamiltonian parameters \cite{046}
(as, for example, an extensive real-space renormalization-group
treatment of the random transverse Ising chain by D.S.Fisher).

One of the interesting subjects in the theory of magnetic
materials is the magnetic relaxation in spin systems and, in
particular, the impurity spin relaxation in spin systems.
Spin-$\frac{1}{2}$ $XY$ chains provide a possibility to rigorously
study the relaxation phenomena in one-dimensional spin models
containing impurities \cite{047,003}. A single impurity may be
assumed in the sense that the interaction with its neighbours is
different in strength ($J^\prime\ne J$). The impurity spin may be
located either at the boundary of the system ($J_1=J^\prime$,
$J_2=J_3=\ldots=J$) or in the bulk (e.g.,
$J_{\frac{N}{2}-1}=J_{\frac{N}{2}}=J^\prime$ and $J_n=J$ for all
other $n$). After the Jordan-Wigner transformation one faces a
chain of tight-binding fermions with the impurity site (in the
sense that the hopping amplitude(s) surrounding this site is (are)
different in value) the energy spectrum of which is well-known (if
$J^\prime <J_c$ the elementary excitation energies form a band
whereas if $J^\prime >J_c$  two states emerge from the band;
moreover, $J_c=\sqrt{2}J$ ($J_c=J$) for the boundary (bulk)
impurity). Nevertheless, the time dependence of the equilibrium
autocorrelation functions $\langle s_n^\alpha (t)s_n^\alpha
\rangle$ is not obvious since they are two-fermion ($zz$) or,
generally speaking, many-fermion ($xx$) quantities. The
time-dependent autocorrelation functions $\langle s_n^\alpha
(t)s_n^\alpha \rangle$ for the impurity spin or for the spins in
its vicinity may exhibit new types of asymptotic behaviour
depending on the relation between $J$ and $J^\prime$ and
temperature. The impurity relaxation may become even more
complicated in the presence of an external (transverse) field.

Spin-$\frac{1}{2}$ $XY$ chains have been used to study the
nonequilibrium properties of quantum systems \cite{048}.
Z.R\'{a}cz with coworkers suggested to consider a nonequilibrium
system imposing a current on a system and investigating the steady
states. Alternatively, we can gain an understanding of the
nonequlibrium properties examining the dynamics of an initial
state (for example, a kink or droplet configuration of $z$ on-site
magnetizations).

Recently
a study of the transport properties
(for example, of the thermal conductivity)
of low-dimensional spin systems
which are significantly determined by magnetic excitations
have attracted much interest
\cite{049}.
A simple case of the isotropic $XY$ chain emerges in such studies
as a milestone
providing reference results
for more sophisticated models.

Finally,
some exotic applications of spin-$\frac{1}{2}$ $XY$ chains
have appeared recently.
For example,
in connection with quantum information processing
the numerical/analytical computations
of entanglement in spin-$\frac{1}{2}$ $XY$ models
on one-di\-men\-si\-o\-nal lattices
of small/infinite number of sites
were carried out
\cite{050}.
A  study of the correlation function
which is called
the emptiness formation probability
(i.e., the probability of the formation of a ferromagnetic string
in the antiferromagnetic ground state)
yields interesting links
between statistical mechanics and number theory.
The case of the isotropic $XY$ chain
provides a valuable background
for calculation of such correlation functions
\cite{051}.

We hope that the present review shows that spin-$\frac{1}{2}$ $XY$
chains still contain quite unexplored properties which deserve to
be discussed. A serious advantage of this type of models is a
possibility to perform statistical mechanical calculations either
rigorously analytically or exactly numerically (considering as
long chains as required to obtain results which pertain to the
thermodynamic limit). On the other hand, it is always desirable to
clarify afterwards the relation of the obtained results to more
realistic models (with the Heisenberg interaction, interchain
interaction, $s>\frac{1}{2}$). Many interesting questions in the
theory of these quantum spin chains still remain open and call for
new efforts.

\newpage

\section{Acknowledgements}

The author thanks Taras Krokhmalskii, Taras Verkholyak and Oles'
Zaburannyi for discussions. This work was partly supported by the
STCU under the project No.~1673. The author is grateful to doctor
Janush Sanotsky who made it possible to prepare the paper in due
time.

\label{last@page}


\begin{thebibliography}{99}

\bibitem{001}
Kenzelmann~M., Coldea~R., Tennant~D.A., Visser~D., Hofmann~M.,
Smeibidl~P., Tylczynski~Z.
// Phys. Rev. B,
2002, vol.~65, 144432.

\bibitem{002}
Farias~G.A., Gon\c{c}alves~L.L.
// Physica status solidi (b),
1987, vol.~139, p.~315.

\bibitem{003}
Stolze~J., N\"{o}ppert~A., M\"{u}ller~G.
// Phys. Rev. B,
1995, vol.~52, p.~4319; Stolze~J., Vogel~M.
// Phys. Rev. B,
2000, vol.~61, p.~4026.

\bibitem{004}
Asakawa~H.
// Physica A,
1996, vol.~233, p.~39.

\bibitem{005}
Young~A.P., Rieger~H.
// Phys. Rev. B,
1996, vol.~53, p.~8486; Young~A.P.
// Phys. Rev. B,
1997, vol.~56, p.~11691; Sachdev~S., Young~A.P.
// Phys. Rev. Lett.,
1997, vol.~78, p.~2220.

\bibitem{006}
Derzhko~O., Krokhmalskii~T.
// Fizika Nizkikh Temperatur (Kharkiv),
1997, vol.~23, p.~721; Phys. Rev. B, 1997, vol.~56, p.~11659;
Physica status solidi (b), 1998, vol.~208, p.~221.

\bibitem{007}
Braeter~H., Kowalski~J.M.
// Physica A,
1977, vol.~87, p.~243; Derzhko~O., Krokhmalskii~T.
// Visnyk L'viv. univ., ser. fiz.,
1993, No. 26, p.~47 (in Ukrainian); Ferroelectrics, 1994, vol.~153, p.~55; J. Magn. Magn. Mater., 1995, vol.~140--144, p.~1623;
Visnyk L'viv. univ., ser. fiz., 1995, No. 27, p.~21 (in
Ukrainian); Ferroelectrics, 1997, vol.~192, p.~21; Derzhko~O.,
Krokhmalskii~T., Verkholyak~T.
// J. Magn. Magn. Mater.,
1996, vol.~157/158, p.~421; Materials Science \& Engineering A,
1997, vol.~226--228, p.~1049; Philosophical Magazine B, 1997, vol.~76, p.~855.

\bibitem{008}
Taylor~J.H.,  M\"{u}ller~G.
// Physica A,
1985, vol.~130, p.~1; Viswanath~V.S., M\"{u}ller~G. The Recursion
Method. Application to Many-body Dynamics. Berlin, Heidelberg,
Springer-Verlag, 1994.

\bibitem{009}
Derzhko~O., Krokhmalskii~T., Stolze~J.
// J. Phys. A,
2000, vol.~33, p.~3063;
Czechoslovak Journal of Physics,
2002, vol.~52, p.~321;
J. Phys. A,
2002, vol.~35, p.~3573.

\bibitem{010}
Derzhko~O., Krokhmalskii~T.
// Physica status solidi (b),
2000, vol.~217, p.~927;
Annalen der Physik (Leipzig),
1999, vol.~8, p.~SI-45.

\bibitem{011}
Barnes~T.
Preprint cond-mat/0204115.

\bibitem{012}
Perk~J.H.H., Capel~H.W.
// Physica A,
1980, vol.~100, p.~1.

\bibitem{013}
Rao~S., Sen~D.
Preprint cond-mat/0005492
(and references therein).

\bibitem{014}
de Lima~J.P., Gon\c{c}alves~L.L.
// Physica A, 2002, vol.~311, p.~458.

\bibitem{015}
Plascak~J.A., Pires~A.S.T., S\'{a} Barreto~F.C.
// Solid State Commun.,
1982, vol.~44, p.~787; Watarai~S., Matsubara~T.
// J. Phys. Soc. Jpn.,
1984, vol.~53, p.~3648; Derzhko~O.V., Levitskii~R.R., Sorokov~S.I.
// Ukrainian Journal of Physics,
1990, vol.~35, p.~1421 (in Ukrainian); Florencio~J., S\'{a}
Barreto~F.C.
// Phys. Rev. B,
1999, vol.~60, p.~9555.

\bibitem{016}
Yukhnovskii~I.R., Levitskii~R.R., Sorokov~S.I., Derzhko~O.V.
// Izv. AN SSSR, ser. fiz.,
1991, vol.~55, p.~481 (in Russian); Sorokov~S.I., Levitskii~R.R.
Thermodynamics and longitudinal dynamical properties of the 1D
Ising model in a transverse field. Preprint of the Institute for
Condensed Matter Physics, ICMP--94--3E, L'viv, 1994, 20~p.;
Levitskii~R.R., Sokolovskii~R.O., Sorokov~S.I.
// Condens. Matter Phys.,
1997, No.~10, p.~67; Levitskii~R.R., Sorokov~S.I., Baran~O.R.
// Condens. Matter Phys.,
2000, vol.~3, p.~515.

\bibitem{017}
Derzhko~O.
// Journal of Physical Studies (L'viv),
2001, vol.~5, p.~49; Derzhko~O., Krokhmalskii~T.
// Acta Physica Polonica B,
2001, vol.~32, p.~3421; J. Magn. Magn. Mater., 2002, vol.~242--245, p.~778; Derzhko~O., Richter~J., Verkholyak~T.
// Acta Physica Polonica B,
2001, vol.~32, p.~3427; Czechoslovak Journal of Physics, 2002,
vol.~52, p.~A41; Derzhko~O., Verkholyak~T., Schmidt~R., Richter~J.
Preprint cond-mat/0207179 (to appear in Physica~A); Nunner~T.S., Kopp~T.
Preprint cond-mat/0210103.

\bibitem{018}
Schulz~H.J.
// Phys. Rev. B,
1986, vol.~34, p.~6372; Kolezhuk~A.K., Mikeska~H.-J.
Preprint cond-mat/0202171.

\bibitem{019}
Tsukada~I., Takeya~J., Masuda~T., Uchinokura~K.
// Phys. Rev. Lett.,
2001, vol.~87, 127203.

\bibitem{020}
Kontorovich~V.M., Tsukernik~V.M.
// Zh. Eksp.~Teor. Fiz.,
1967, vol.~52, p.~1446 (in Russian); Siskens~Th., Capel~H.W.,
Gaemers~K.J.F.
// Physica A,
1975, vol.~79, p.~259; Siskens~Th., Capel~H.W. // Physica A, 1975,
vol.~79, p.~296; Zvyagin~A.A.
// Phys. Lett. A,
1991, vol.~158, p.~333; Daniel~M., Amuda~R.
// Phys. Rev. B,
1996, vol.~53, p.~R2930; Gottlieb~D., R\"{o}ssler~J.
// Phys. Rev. B,
1999, vol.~60, p.~9232; Pires~A.S.T.
// J. Magn. Magn. Mater.,
2001, vol.~223, p.~304.

\bibitem{021}
Oshikawa~M., Affleck~I.
// Phys. Rev. Lett.,
1997, vol.~79, p.~2883; Aristov~D.N., Maleyev~S.V.
// Phys. Rev. B,
2000, vol.~62, p.~R751; Derzhko~O., Richter~J., Zaburannyi~O.
// J. Phys.: Condens. Matter,
2000, vol.~12, p.~8661.

\bibitem{022}
Derzhko~O.V., Levitskii~R.R., Moina~A.Ph.
// Condens. Matter Phys.,
1993, No. 1, p.~115 (in Ukrainian); Derzhko~O.V., Moina~A.Ph.
// Condens. Matter Phys.,
1994, No. 3, p.~3; Ferroelectrics, 1994, vol.~153, p.~49; Physica
status solidi (b), 1996, vol.~196, p.~237.

\bibitem{023}
Bocquet~M., Essler~F.H.L., Tsvelik~A.M., Gogolin~A.O.
Preprint cond-mat/0102138.

\bibitem{024}
Lloyd~P.
// J. Phys. C,
1969, vol.~2, p.~1717; Nishimori~H.
// Phys. Lett. A,
1984, vol.~100, p.~239; Derzhko~O., Verkholyak~T.
// Physica status solidi (b),
1997, vol.~200, p.~255; Materials Science \& Engineering A, 1997,
vol.~226--228, p.~745; Fizika Nizkikh Temperatur (Kharkiv), 1997,
vol.~23, p.~977.

\bibitem{025}
John~W., Schreiber~J.
// Physica status solidi (b),
1974, vol.~66, p.~193; Richter~J., Handrich~K., Schreiber~J.
// Physica status solidi (b),
1975, vol.~68, p.~K61; Richter~J., Schreiber~J., Handrich~K.
// Physica status solidi (b),
1976, vol.~74, p.~K125; Richter~J.
// Physica status solidi (b),
1978, vol.~87, p.~K89.

\bibitem{026}
Derzhko~O., Richter~J.
// Phys. Lett. A,
1996, vol.~222, p.~348;
Phys. Rev. B,
1997, vol.~55, p.~14298;
Phys. Rev. B,
1999, vol.~59, p.~100.

\bibitem{027}
Derzhko~O., Richter~J., Derzhko~V.
// Annalen der Physik (Leipzig),
1999, vol.~8, p.~SI-49.

\bibitem{028}
Gon\c{c}alves~L.L., Vieira~A.P.
// J. Magn. Magn. Mater.,
1998, vol.~177/181, p.~79.

\bibitem{029}
Derzhko~O., Krokhmalskii~T.
// Journal of Physical Studies (L'viv),
1998, vol.~2, p.~263; Derzhko~O., Krokhmalskii~T., Zaburannyi~O.
// Condens. Matter Phys.,
1999, vol.~2, p.~339.

\bibitem{030}
Nieuwenhuizen~T.M.
// J.~Phys. A,
1984, vol.~17, p.~1111; Nieuwenhuizen~T.M., Luck~J.M.
// J.~Phys. A,
1986, vol.~19, p.~1207; Luck~J.M., Nieuwenhuizen~T.M.
// J.~Phys. A,
1989, vol.~22, p.~2151; Funke~M., Nieuwenhuizen~T.M., Trimper~S.
// J.~Phys. A,
1989, vol.~22, p.~5097; Luck~J.M., Funke~M., Nieuwenhuizen~T.M.
// J.~Phys. A,
1991, vol.~24, p.~4155.

\bibitem{031}
de~Lima~J.P., Gon\c{c}alves~L.L.
// J. Magn. Magn. Mater.,
1995, vol.~140--144, p.~1606; J. Magn. Magn. Mater., 1999, vol.~206, p.~135; Barbosa Filho~F.F., de~Lima~J.P., Gon\c{c}alves~L.L.
// J. Magn. Magn. Mater.,
2001, vol.~226--230, p.~638; Tong~P., Zhong~M.
// Physica B,
2001, vol.~304, p.~91; Stre\v{c}ka~J., Ja\v{s}\v{c}ur~M.
// Czechoslovak Journal of Physics,
2002, vol.~52, p.~A37.

\bibitem{032}
Derzhko~O.
// Fizika Nizkikh Temperatur (Kharkiv),
1999, vol.~25, p.~575; Derzhko~O., Richter~J., Zaburannyi~O.
// Phys. Lett. A,
1999, vol.~262, p.~217;
Acta Physica Polonica A,
2000, vol.~97, p.~931;
Physica A,
2000, vol.~282, p.~495;
J. Magn. Magn. Mater.,
2000, vol.~222, p.~207.

\bibitem{033}
Derzhko~O.
// J. Phys. A,
2000, vol.~33, p.~8627; Derzhko~O., Zaburannyi~O.
// Ukrainian Journal of Physics,
2002, vol.~47, p.~599 (in Ukrainian); Derzhko~O., Richter~J.,
Zaburannyi~O.
// J. Magn. Magn. Mater.,
2002, vol.~242--245, p.~778; Derzhko~O., Richter~J.,
Krokhmalskii~T., Zaburannyi~O.
// Phys. Rev. B,
2002, vol.~66, p.~144401.

\bibitem{034}
Derzhko~O.
// Czechoslovak Journal of Physics,
2002, vol.~52, p.~A277.

\bibitem{035}
Perk~J.H.H., Capel~H.W., Zuilhof~M.J., Siskens~Th.J.
// Physica A,
1975, vol.~81, p.~319.

\bibitem{036}
Oshikawa~M., Yamanaka~M., Affleck~I.
// Phys. Rev. Lett.,
1997, vol.~78, p.~1984.

\bibitem{037}
Sachdev~S. Quantum Phase Transitions. New York, Cambridge
University Press, 1999.

\bibitem{038}
Idogaki~T., Rikitoku~M., Tucker~J.W.
// J. Magn. Magn. Mater.,
1996, vol.~152, p.~311; Derzhko~O., Zaburannyi~O.
// Journal of Physical Studies (L'viv),
1998, vol.~2, p.~128; Derzhko~O., Zaburannyi~O., Tucker~J.W.
// J. Magn. Magn. Mater.,
1998, vol.~186, p.~188.

\bibitem{039}
Drell~S.D., Weinstein~M., Yankielowicz~S.
// Phys. Rev. D,
1977, vol.~16, p.~1769.

\bibitem{040}
Cabra~D.C., Grynberg~M.D., Honecker~A., Pujol~P.
Preprint cond-mat/0010376 (and references therein);
Grynberg~M.D., Cabra~D.C., Arlego~M.
// Phys. Rev.~B, 2001, vol.~64, p.~134419;
Derzhko~O.
// Ukrainian Journal of Physics,
2001, vol.~46, p.~762.

\bibitem{041}
Pincus~P.
// Solid State Commun.,
1971, vol.~9, p.~1971; Beni~G., Pincus~P.
// J.~Chem.~Phys.,
1972, vol.~57, p.~3531; Beni~G.
// J. Chem. Phys.,
1973, vol.~58, p.~3200.

\bibitem{042}
Sil~S.
// J. Phys.: Condens. Matter,
1998, vol.~10, p.~8851.

\bibitem{043}
Boucher~J.P., Regnault~L.P.
// J. Phys. I France,
1996, vol.~6, p.~1939.

\bibitem{044}
Derzhko~O., Krokhmalskii~T.
// Ferroelectrics,
2001, vol.~250, p.~397.

\bibitem{045}
Okamoto~K.
// Solid State Commun.,
1992, vol.~83, p.~1039.

\bibitem{046}
Pfeuty~P.
// Phys. Lett. A,
1979, vol.~72, p.~245; Luck~J.M.
// Journal of Statistical Physics,
1993, vol.~72, p.~417; Fisher~D.S.
// Phys. Rev. B,
1995, vol.~51, p.~6411; McKenzie~R.H.
// Phys. Rev. Lett.,
1996, vol.~77, p.~4804; Igl\'{o}i F., Turban~L., Karevski~D.,
Szalma~F.
// Phys. Rev. B,
1997, vol.~56, p.~11031; Henelius~P., Girvin~S.M.
// Phys. Rev. B,
1998, vol.~57, p.~11457; Hermisson~J., Grimm~U., Baake~M.
Preprint cond-mat/9706106;
Hermisson~J.
Preprint cond-mat/9808238.

\bibitem{047}
Tjon~J.A.
// Phys. Rev. B,
1970, vol.~2, p.~2411.

\bibitem{048}
Antal~T., R\'{a}cz~Z., Sasv\'{a}ri~L.
// Phys. Rev. Lett.,
1997, vol.~78, p.~167; Antal~T., R\'{a}cz~Z., R\'{a}kos~A.,
Sch\"{u}tz~G.M.
// Phys. Rev. E,
1998, vol.~57, p.~5184; Antal~T., R\'{a}cz~Z., R\'{a}kos~A.,
Sch\"{u}tz~G.M.
// Phys. Rev. E,
1999, vol.~59, p.~4912; R\'{a}cz~Z.
// Journal of Statistical Physics,
2000, vol.~101, p.~273; Ogata~Y.
// Phys. Rev. E,
2002, vol.~66, 016135; Karevski~D.
// Eur. Phys. J. B,
2002, vol.~27, p.~147; Berim~G.O., Cabrera~G.G.
// Physica A,
1997, vol.~238, p.~211; Berim~G.O., Berim~S., Cabrera~G.G.
// Phys. Rev. B, 2002, vol.~66, p.~094401.

\bibitem{049}
Heidrich-Meisner~F., Honecker~A., Cabra~D.C., Brenig~W.
// Phys. Rev. B, 2002, vol. 66, p.~140406(R).

\bibitem{050}
Fu~H., Solomon~A.I., Wang~X.
// J. Phys. A,
2002, vol.~35, p.~4293; Osborne~T.J., Nielsen~M.A.
Preprint quant-ph/0202162;
Bose~I., Chattopadhyay~E.
Preprint cond-mat/0208011.

\bibitem{051}
Shiroishi~M., Takahashi~M., Nishiyama~Y.
Preprint cond-mat/0106062;
Boos~H.E., Korepin~V.E., Nishiyama~Y., Shiroishi~M.
Preprint cond-mat/0202346;
Abanov~A.G., Korepin~V.E.
Preprint cond-mat/0206353.

\end{thebibliography}
\end{document}